# HIGH QUALITY AUDIO CODING WITH MDCTNET

*Grant Davidson¹, Mark Vinton¹, Per Ekstrand², Cong Zhou¹\*, Lars Villemoes², and Lie Lu¹*

¹Dolby Laboratories, Inc., 1275 Market Street, San Francisco, CA, 94103, USA
²Dolby Sweden AB, Gävlegatan 12A, 113 30, Stockholm, Sweden

## ABSTRACT

We propose a neural audio generative model, MDCTNet, operating in the perceptually weighted domain of an adaptive modified discrete cosine transform (MDCT). The architecture of the model captures correlations in both time and frequency directions with recurrent layers (RNNs). An audio coding system is obtained by training MDCTNet on a diverse set of fullband monophonic audio signals at 48 kHz sampling, conditioned by a perceptual audio encoder. In a subjective listening test with ten excerpts chosen to be balanced across content types, yet stressful for both codecs, the mean performance of the proposed system for 24 kb/s variable bitrate (VBR) is similar to that of Opus at twice the bitrate.

## 1. INTRODUCTION

The rapid evolution of generative modeling has changed our expectation on the efficiency of wideband speech coding [1-4]. For general audio coding, the main tracks of research have been to improve the outputs of traditional codecs [5-10] or to train a neural net end-to-end as an autoencoder for regression according to a perceptually motivated loss function [11]. The performance gains have been moderate with methods solely addressing redundancy and irrelevancy, although these aspects are sufficient for optimality in a philosophical sense [12]. The introduction of an adversarial loss component in [13] has confirmed the benefits of being able to synthesize plausible signals according to the dataset, especially at very low bitrates. It has also been shown [14] that wideband coding of speech and piano signals can be improved by training a generative model conditioned on a perceptual audio codec, but it has remained a challenge to generalize this method to arbitrary fullband (0-20 kHz) signal categories and demonstrate significant improvement relative to state-of-the-art conventional codecs.

In this paper we present a novel fullband audio codec based on a combination of perceptual audio coding and deep learning principles that offers robust subjective performance across a diverse set of audio signals. The encoder is based on a block-switched MDCT [15] with a quantization and coding module driven by an advanced perceptual model. A low-rate perceptual domain signal representation is conveyed to the decoder. In the decoder, we integrate a traditional IMDCT-based decoding module with a generative model conditioned on the received perceptual-domain signals. This model, called MDCTNet, is based on an architecture that exploits signal redundancy across both frequency and time. The model has been trained on monophonic audio at rates from 20-32 kb/s VBR. A subjective assessment indicated that at 24 kb/s the deep codec offers an overall-mean performance equal to the Opus switched speech-transform codec [16] operating at twice the bitrate. Notably, in this test the deep codec also offered more consistent subjective quality across content types.

The remainder of this paper is organized as follows. Section 2 motivates the use of perceptual domain signals for deep audio processing and describes the core transform encoder/decoder structures. In Section 3 we present details of the MDCTNet architecture. Our objective and subjective experiment results are presented in Section 4, with concluding remarks in Section 5.

## 2. ENCODER AND DECODER ARCHITECTURE

The overall architecture of the transform domain encoder and generative neural network-based decoder is shown in Figure 1. The encoder process is similar to a typical transform domain codec like AAC [17]. The input PCM waveform is first converted to the frequency domain using an MDCT. To ensure high performance on both stationary and non-stationary signals, the MDCT transform size can be switched between 768 samples for stationary content and a shorter 192 samples for non-stationary content. The block-switch is the same mechanism used in AAC with bridge windows between the long and short transforms. The window sequence for the MDCT is transmitted to the decoder as side information. In parallel to the MDCT the signal masking threshold is calculated with a psychoacoustic model. The psychoacoustic model we use in the encoder is based on a gammatone filter bank similar to that proposed in [18]. The masking threshold is then used to construct a perceptual envelope that is used to

---

weight the MDCT spectrum in sub-bands. We refer to the weighted MDCT spectrum as the perceptual domain as the signal magnitude of the weighted MDCT spectrum in each time block and frequency sub-band is proportional to perceptual relevance of the signal. The perceptual envelope is quantized and entropy encoded prior to transmission to the decoder as side information. To reduce side-information cost, the perceptual envelope can be shared over multiple blocks for stationary signals. The final stage of the encoder process is the quantization and entropy encoding of the MDCT coefficients. The overall bitrate of the encoded signal is controlled by a file based VBR technique such that the target bitrate is averaged over the entire encoded file.

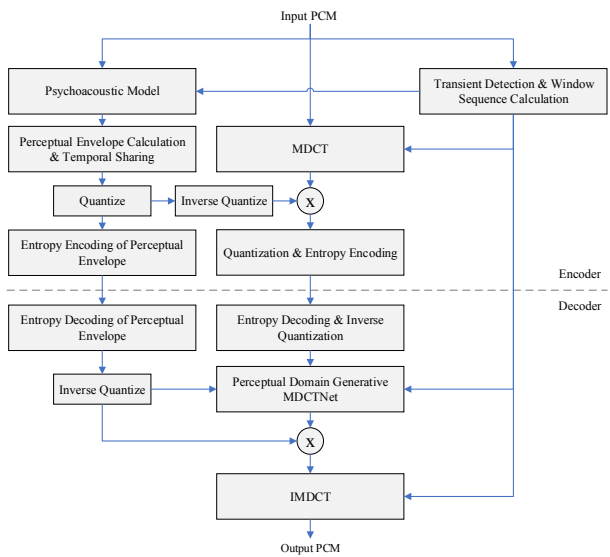

**Figure 1.** Schematic of the encoder and decoder structure deploying a generative neural network in the decoder

With reference to the decoder section of Figure 1, both the MDCT spectrum and the perceptual envelope are first entropy decoded and inverse quantized. The inverse quantized transform coefficients and perceptual envelope are combined with the window sequence information to condition the generative neural network. The architecture of the perceptual domain generative MDCTNet is described in detail in Section 3. However, an important feature of the overall system architecture is that the neural network is trained and inferenced in the perceptual domain. Since the perceptual domain has a significantly reduced dynamic range relative to the raw unweighted MDCT spectrum, we found that the training procedure on a large diverse dataset became tractable and the network was more robust to different content types. Indeed, many of the items we used for the subjective testing presented in section 4.3 are from content types that are not well represented in our training set. The output of the generative MDCTNet is re-normalized by the perceptual envelope in sub-bands to convert the perceptual domain signal back to the MDCT domain. The final stage of the decoder is the block-switched IMDCT that converts the frequency domain signal back to the time domain.

## 3. NETWORK ARCHITECTURE

The task of the MDCTNet generative model is to synthesize plausible signals conditioned on the information from the encoder. As discussed in [14], this encourages a perceptual noise shaping as designed in the encoder, but in contrast to [14] the MDCTNet operates directly in the perceptual domain. To capture correlations in time and frequency, the generative model makes use of recurrent layers (RNNs) in both time and frequency directions. Similar ideas have been considered for speech coding [19], audio synthesis [20], and music source separation [21].

For block switching, the short MDCT coefficient vectors of 192 lines are individually repeated by a factor four in the frequency direction to 768 lines. Likewise, the perceptual envelopes containing either 19 or 44 bands are mapped to a 768-line representation. Hence, the generative model receives a sequence of frames having an equal number of MDCT coefficients and perceptual domain envelope parameters, but with an adaptive time grid defined by the transform window sequence. This sequence is presented to the network as a one-hot vector of 4 channels representing the 4 different window types.

### 3.1. Predictor high level structure

The time direction RNN predicts coefficients grouped into a target band based on the context from all previous times ($< t$) of the same band (red traces). The cross-band predictor computes coefficients in a target band based on the past neighboring bands [19] (green traces). The frequency direction RNN predicts coefficients in a target band $b$ based on the previous lower frequency bands ($< b$) at the current time $t$ (dark blue trace). The high-level predictor structure is illustrated in Figure 2.

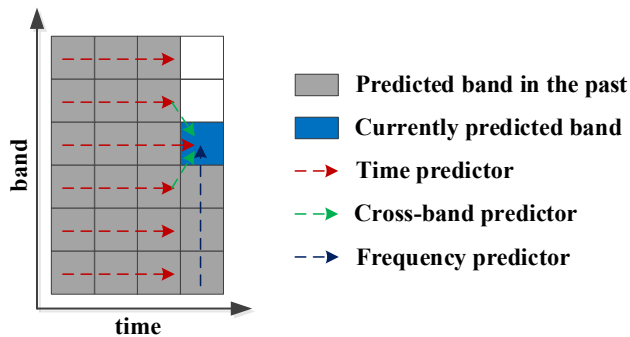

**Figure 2.** High-level diagram of the predictor structure

### 3.2. Model architecture

The MDCTNet is a generative model operating on transform domain coefficients. The joint probability density of a vector of MDCT coefficients $X$ is factorized as a product of conditional probabilities, as shown in Equation 1.

$$p(X|C) = \prod_b \prod_t p(X_{t,b} \mid X_{0...t-1,b-N...b+N}, X_{t,0...b-1}, c) \quad (1)$$

where $X_{t,b}$ denotes the coefficients at time $t$ in band $b$, $X_{0...t-1,b-N...b+N}$ represents a set of coefficients of past target and neighboring bands, and $X_{t,0...b-1}$ is the previous (lower) band coefficients of time $t$.

The model is also conditioned on data $c$ from the waveform coder. In the proposed model $c$ is context from a conditioning network having quantized coefficients $\widetilde{X}_{t,b}$, perceptual domain envelope $e_{t,b}$, and the windows sequence $w_t$ as input.

With reference to Figure 3, we will now present the model architecture. The previous coefficient vector $X_{t-1}$ is first grouped into frequency bands by a convolution layer (A), resulting in a latent of the internal dimension (1024). In parallel to A, the conditioning network processes $\widetilde{X}_t$ and $e_t$ using a similar layer (B) but employing larger horizontal kernel size to increase the receptive field to future samples ($n = 6$ in the proposed model). The one-hot encoded $w_t$ is processed by another convolutional layer C, and added to the output of layer B. The summed result is split into two tensors, where one of them is directly added to the output tensor from layer A, which constitutes the input to the RNN running in the time direction.

The time direction RNN is a dual layer Gated Recurrent Unit (GRU), designed to capture temporal dependencies. The RNN processes $B$ band sequences separately using shared weights but having individually learned initial hidden states for each band. The hidden states from the time direction RNN are subsequently concatenated to the other half of the output from the conditioning network and subjected to a subsequent convolutional layer (D). This layer provides the cross-band prediction (Figure 2), and the output from this layer denotes the final time domain context.

To prepare the coefficients $X_t$ for prediction in the frequency direction, the coefficients in each band $b$ is projected by a convolutional layer E. The input $X_t$ to layer E is first organized so that each output band $b$ has the coefficients from previous band $b - 1$ as input. Hence, the coefficients of band $b - 1$ are projected to band $b$ using a $1 \times 1$ kernel.

The context from layer E is added to the time domain context, and this sum is subjected to the RNN running in the frequency direction. The frequency direction RNN is likewise a dual-layer GRU, providing stateful prediction to capture patterns in the frequency domain, such as harmonic relations. The initial hidden states of the RNN are also learnable.

Finally, the hidden states from the frequency direction RNN are provided to a Multi-layer Perceptron (MLP), consisting of two convolutional layers preceded by Rectified Linear Unit (ReLU) activations. The output from the MLP consists of a group of Laplacian distribution model parameters $\mu$ (location) and $s$ (scale).

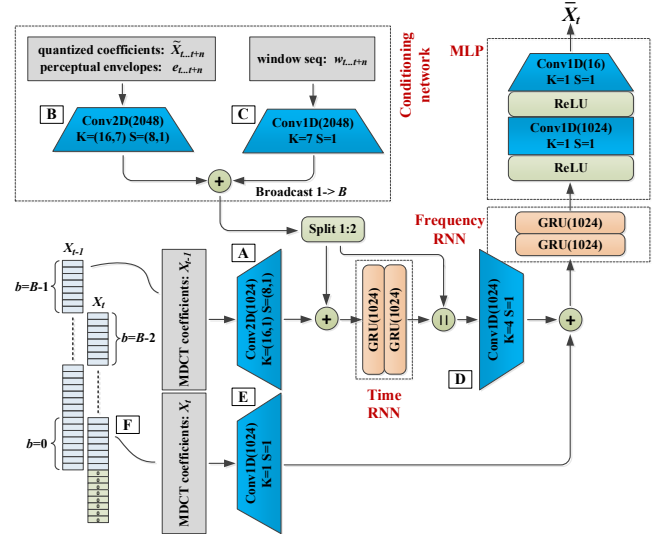

**Figure 3.** MDCTNet model architecture details.

In inferencing mode, i.e., *self-generation*, random samples of the Laplacian distribution according to the model parameters are used to create MDCT coefficients $\overline{X}_t(b)$ which are sequentially fed back to the input buffer (F) and used as past samples. In the training phase, the ground truth MDCT coefficients are used as past samples during prediction (this is known as *teacher forcing*). The training objective stochastically minimizes the Laplacian distribution-based negative log-likelihood of the target (ground truth) MDCT coefficients $y$ given the model output parameters, as shown in Equation 2.

$$NLL(\mu, s, y) = \log(2s) + \frac{|\mu - y|}{s} \quad (2)$$

## 4. EVALUATION RESULTS

### 4.1. Model training

With the configuration shown in Figure 3, our model has approximately 35M parameters. For model training we utilized a diverse collection of 720 hours of open and internal audio datasets, including a partially-balanced set of 18 different music genres, speech, and individual instruments. The audio, stored as 48 kHz sampled files, was encoded into bitstreams at a rate of 24 kb/s VBR. During training, we used a batch size of 32 with each bitstream randomly cropped to 1 second for each mini-batch. We used the ADAM optimizer [22] ($\beta_1 = 0.9$, $\beta_2 = 0.999$, and $\varepsilon = $ 1e-8) with an initial learning rate of 0.0002. The learning rate was reduced by one-half when validation loss did not decrease after three consecutive epochs.

**4.2. Objective Evaluation**

To illustrate the proficiency of the MDCTNet model to generate plausible signals from a sparse perceptual-domain representation, Figure 4 presents spectrograms of original and 24 kb/s coded violin signals. The middle spectrogram represents the output of the core codec, (which bypasses MDCTNet in Figure 2), showing numerous spectral holes and missing harmonics. In the bottom spectrogram the missing information has been plausibly generated with harmonics continuous in time and occurring with a frequency spacing matching the original signal. From the figure we observe that the network is simultaneously performing spectral hole filling and adaptive-crossover frequency bandwidth extension.

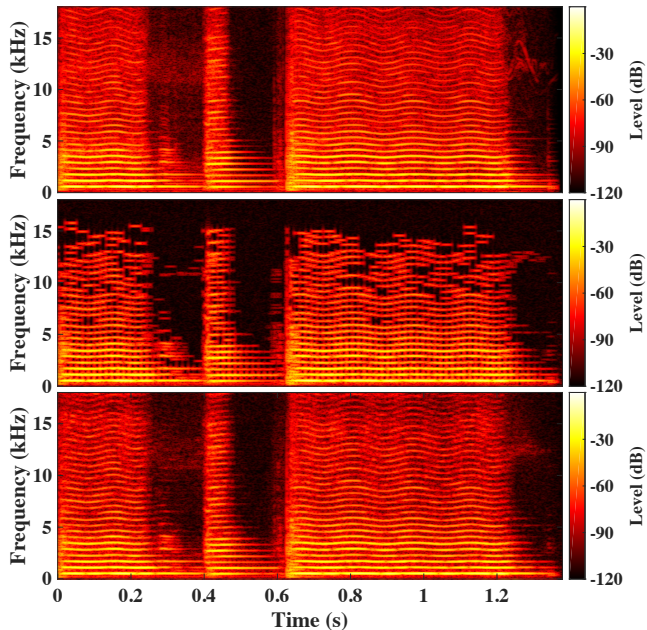

**Figure 4.** 0–18 kHz spectrogram comparison of original violin signal with vibrato (top), 24 kb/s core codec decoded signal (middle), and 24 kb/s MDCTNet decoded signal (bottom).

**4.3. Subjective Evaluation**

We completed a subjective evaluation of our deep codec at 24 kb/s compared to Opus at 48 kb/s, both mono VBR. In accordance with ITU-R recommendations, we applied the MUSHRA methodology for evaluating intermediate quality impairments [23]. A panel of 18 experienced listeners assessed, over high-quality headphones, the basic audio quality based on any and all detected differences between the known reference and the systems under test. Each trial included the two codecs, a hidden reference, and 3.5 and 7 kHz lowpass anchors. The average bitrate of each encoded file was controlled to within ±5% of the target bitrate.

A search for critical audio was completed by a panel of four expert listeners over a set of 51 items representing typical broadcast program material. Ten items were selected for testing that 1) were most stressful to each codec under test, and 2) represented balance across various content types. The final set of ten items included clean speech, voice mixed with music, solo vocals, applause, individual instruments, and musical selections. None of the test items were included in the MDCTNet training and validation datasets.

The subjective evaluation results are presented in Figure 5. The ITU-R recommended listener post-screening rules were applied. We found that the overall mean performance of both codecs was rated near the bottom of the "Excellent" range. A pairwise difference test revealed that the overall mean performance of the codecs must be considered statistically identical (95% confidence). Now considering the per-item mean ratings, we see that the worst-case item for MDCTNet (Applause) was rated as "Good," while the worst-case for Opus (Muted Trumpet) was rated near the top of the "Fair" range. MDCTNet performed best on items containing tonal signal components, and Opus, with its integrated voice codec, performed best on clean speech and voice mixed with music. In general, the performance pattern of the MDCTNet was more consistent across items.

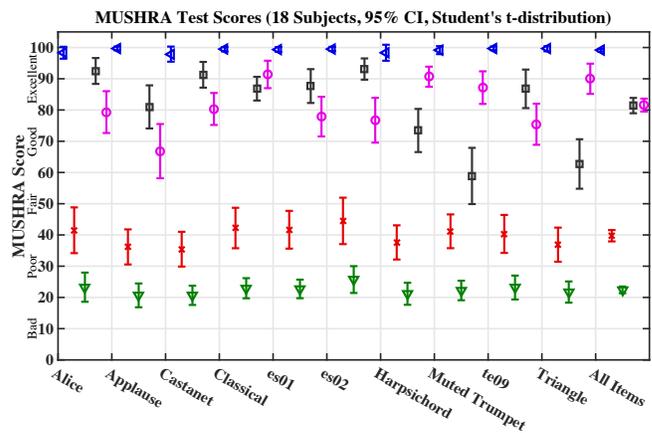

**Figure 5.** Results of the BS.1534-3 listening test comparing the MDCTNet at 24 kb/s VBR and Opus 48 kb/s VBR.

**5. CONCLUSIONS**

In this paper we presented a deep codec including MDCTNet as a neural generative model that generalizes to arbitrary fullband audio signals and demonstrates significant performance gains relative to a traditional audio codec. We found that training the model on perceptual domain signals derived from a diverse and balanced dataset enabled the network to efficiently capture the essence of general audio content. Although other work has demonstrated good-to-excellent speech/audio quality when operating at lower sample rates and/or when trained on narrower signal classes, to our knowledge this deep codec is the first to offer compelling performance on typical 48 kHz broadcast content. In the future we believe it is worthwhile to explore the quality-rate tradeoff by quantifying performance at different bitrates.